\documentclass[groupedaddress,
 reprint,
 amsmath,amssymb,
 nofootinbib,
 superscriptaddress,
 aps,
 prd,
]{revtex4-1}

\usepackage{dcolumn}
\usepackage{bm}
\usepackage{hyperref}
\usepackage[utf8]{inputenc}
\usepackage[capitalise]{cleveref}
\usepackage{aas_macros}
\usepackage[normalem]{ulem}
\usepackage{xcolor}
\usepackage{csquotes}
\usepackage{graphicx}

\usepackage[makeroom]{cancel}
\usepackage{printlen}

\usepackage{mathtools}
\usepackage{physics}

\usepackage{amsmath}

\makeatletter
\newcommand*\rel@kern[1]{\kern#1\dimexpr\macc@kerna}
\newcommand*\widebar[1]{%
  \begingroup
  \def\mathaccent##1##2{%
    \rel@kern{0.8}%
    \overline{\rel@kern{-0.8}\macc@nucleus\rel@kern{0.2}}%
    \rel@kern{-0.2}%
  }%
  \macc@depth\@ne
  \let\math@bgroup\@empty \let\math@egroup\macc@set@skewchar
  \mathsurround\z@ \frozen@everymath{\mathgroup\macc@group\relax}%
  \macc@set@skewchar\relax
  \let\mathaccentV\macc@nested@a
  \macc@nested@a\relax111{#1}%
  \endgroup
}
\makeatother
\newcommand{\mean}[1]{\widebar{#1}}

\begin{document}

\title{Evidence for axion miniclusters with an increased central density}

\author{Benedikt Eggemeier}
\email{benedikt.eggemeier@phys.uni-goettingen.de}
\affiliation{
 Institut f\"ur Astrophysik, Georg-August-Universit\"at G\"ottingen, D-37077 G\"ottingen, Germany
}

\author{Ananthu Krishnan Anilkumar}
\email{a.anilkumar@stud.uni-goettingen.de}
\affiliation{
 Institut f\"ur Astrophysik, Georg-August-Universit\"at G\"ottingen, D-37077 G\"ottingen, Germany
}

\author{Klaus Dolag}
\email{kdolag@mpa-garching.mpg.de}
\affiliation{Universitäts-Sternwarte, Fakultät für Physik, Ludwig-Maximilians-Universität München, Scheinerstr.1, 81679 München, Germany }
\affiliation{Max-Planck-Institut für Astrophysik, Karl-Schwarzschild-Straße 1, 85741 Garching, Germany}

\date{\today}

\begin{abstract} 
We identify axion miniclusters collapsing in the radiation-dominated era and follow them to redshift $z=99$ with N-body simulations. 
We find that the majority of the densest miniclusters end up in the center of larger minicluster halos at late times. 
Soon after their formation, the miniclusters exhibit NFW profiles but they subsequently develop a steeper inner slope approaching $\rho\sim r^{-2}$ on small scales.
Using the so far most highly resolved axion structure formation simulation with $2048^3$ particles we examine the structure of previously studied minicluster halos. 
While the density profiles of their subhalos are NFW-like we confirm that a modified NFW profile with a steeper inner slope provides a better description for minicluster halos with masses above $\sim 10^{-12}\,M_\odot$. 
We show that miniclusters with a higher central density might be in contrast to pure NFW halos dense enough to induce gravitational microlensing. Likewise, more compact minicluster halos will have immediate implications for direct and indirect axion detection.  
\end{abstract}

\maketitle

\section{Introduction}

The QCD axion is a hypothetical particle and results from a spontaneously broken global $U(1)$ symmetry that was introduced by Peccei and Quinn to provide a solution to the so-called strong CP problem~\cite{Weinberg1978,Wilczek1978,Kim1979,Shifman1980,Dine1981,Preskill1983,PQ1,PQ2}.
Since the axion is stable and only very weakly coupled to the Standard Model it is an attractive dark matter candidate~\cite{Marsh_review}. 
The point in time of symmetry-breaking is crucial for the subsequent evolution of the axion and thus affects the phenomenology of axion dark matter.

In the post-inflationary symmetry-breaking scenario the early axion field is characterized by small-scale inhomogeneities which further leads to the formation of a cosmic string network~\cite{Kibble1976}. 
When the axion mass becomes relevant compared to the expansion rate of the universe, domain walls start to build between the axion strings. This initializes the collapse of the string network producing both relativistic and nonrelativistic axions. 
As highly nonlinear processes are involved, the generation of axions from the decay of axion strings can only be studied numerically which is done via cosmological lattice simulations~\cite{Fleury2015,Klaer2017_1,Klaer2017_2,Gorghetto2018,Vaquero2019,Buschmann2019,Gorghetto2020,Buschmann2021,OHare2021,Saikawa2024}. This is a highly non-trivial and challenging task that allows for the prediction of axion model parameters and the dark matter abundance. While there are some simulation scheme-dependent systematic uncertainties concerning the axion mass (see for example Ref.~\cite{Hoof2021} for a discussion), the occurrence of large overdensities in the axion field as a consequence of the collapse of axion strings is a firm prediction.  

Eventually, gravity becomes the dominant force and the axion overdensities collapse into gravitationally bound axion \textit{miniclusters} (MCs)~\cite{Hogan1988,Kolb1993,Kolb1993_2,Kolb1994,Kolb1995}. 
In contrast to the usual structure formation scenarios with cold dark matter (CDM), axion MCs form already during the radiation-dominated era of cosmic history with typical masses of $M\sim 10^{-12}\,M_\odot$ and merge hierarchically into larger structures known as axion \textit{minicluster halos}~(MCHs)~\cite{fairbairn2018,Eggemeier2019_2}.
Since MCs are particularly dense objects it is expected that a large fraction survives the merging processes. Hence, axion MCs should exist as dark matter substructures within galaxy-sized halos at present if they are not tidally disrupted by encounters with stars~\cite{Tinyakov2016,Dokuchaev2017,Kavanagh2020,Shen2022,Dandoy2022,OHare2023}. 

Their presence in Milky Way-like dark matter halos is of significance for axion searches relying on direct and indirect detection techniques. If a large fraction of dark matter is bound in axion MCs, opportunities for indirect axion detection arise from gravitational microlensing~\cite{Kolb1995,Fairbairn2017,fairbairn2018} and transient radio signals originating from collisions of axion MCs with the magnetospheres of neutron stars~\cite{Tkachev2014,Pshirkov2016,Edwards2020}. 
Conversely, this decreases the chances of a direct axion detection in haloscope experiments significantly as an encounter of an MC with the Earth is presumably a rare event taking place only once every $10^5$ years~\cite{Tinyakov2016,Irastorza2018}. 
Nevertheless, the probability of detecting axions directly is enhanced by the possible tidal disruption of MCs and MCHs leading to tidal streams that cover a larger volume with axions. 

To make reliable quantitative predictions about the distribution of axion MCs and their structure, the gravitational growth and collapse of axion overdensities need to be studied numerically. Starting from initial conditions produced by lattice simulations of the early axion field evolution using the methods from Ref.~\cite{Vaquero2019}, the first N-body simulations addressing the formation of axion MCHs were presented in Ref.~\cite{Eggemeier2019_2}. 
They found that at their final redshift of $z=99$ roughly $75\%$ of axions are gravitationally bound in MCHs covering a mass range from $10^{-15}\,M_\odot$ to $10^{-9}\,M_\odot$ and they analyzed the structure of their most massive MCHs. 
Focussing on the \enquote{empty} space between axion MCHs (so-called axion minivoids) it was discovered that the axion density within the voids is only $\sim 10\%$ of the expected galactic dark matter density~\cite{Pierobon2022_void} which has immediate consequences for axion haloscope experiments.\footnote{Note that the average void density obtained in Ref.~\cite{Pierobon2022_void} at \mbox{$z=99$} cannot be directly extrapolated to the present as the tidal disruption of MCHs in the galaxy needs to be considered~\cite{Kavanagh2020,OHare2023}.} 
Furthermore, the Peak-Patch algorithm~\cite{Stein2018} was applied to the same initial axion density field~\cite{Ellis2020,Ellis2022} to identify particularly dense axion MCs. It was realized that some of them might be sufficiently dense to generate observable gravitational microlensing events if the axion mass falls within the mass range $0.2\,\mathrm{meV}\lesssim m_a \lesssim 3\,\mathrm{meV}$~\cite{Ellis2022}.
Considering a purely white-noise power spectrum corresponding to the largest scales of the isocurvature initial power spectrum of Ref.~\cite{Eggemeier2019_2}, N-body simulations were performed with much larger box sizes reaching significantly lower redshifts~\cite{Xiao2021}. While not being able to resolve the formation of axion MCs, the formation of larger and more massive MCHs with typical masses of $\sim 10^{-7}\,M_\odot$ at $z=19$ was observed. Extrapolating the results to redshift $z=0$, MCHs with masses of up to $10^{-4}\,M_\odot$ were predicted. 
It was shown that pulsar timing and microlensing probes may be sensitive to these axion substructures if they are not disrupted.  

There might be significant differences between axion MCs and MCHs (see \cref{sec:MCs_MCHs} for details) that have not been examined in numerical studies so far. 
The radial density profile of an axion MC(H) is crucial to determining if the object survives stellar encounters~\cite{Kavanagh2020} and if it can generate signatures that are observable with indirect detection techniques~\cite{Edwards2020,Ellis2022}.  
Until now, the density profiles of MCHs have been studied at the redshift $z_\mathrm{eq}$ of matter-radiation equality and at $z=99$~\cite{Eggemeier2019_2,Ellis2022,Pierobon2023}. The spatial resolution of the underlying N-body simulations decreases with time, so the MCH density profiles could not be properly resolved on small radial scales. While higher-mass MCHs were found to be in good agreement with Navarro-Frenk-White~(NFW) density profiles~\cite{NFW}, lower-mass MCHs tend to be better described by single power-law profiles~\cite{Eggemeier2019_2,Ellis2022}. 
However, it is uncertain if this difference between lower- and higher-mass objects arises from a lack of spatial resolution. 
Moreover, the formation of MCs collapsing during the radiation-dominated epoch and their evolution has not been studied yet. It is conceivable that the density profiles of early-forming MCs differ from those of larger MCHs originating from the hierarchical merging of MCs.

In this work, we shed light on the structure of axion MCs at the time of their formation, how they evolve into larger MCHs, and how they affect the structure of MCHs. Thus, we aim to contribute to a more comprehensive understanding of axion dark matter structure formation. 
Based on the N-body simulations of Ref.~\cite{Eggemeier2019_2} with $1024^3$ particles, we identify dense MCs forming during radiation domination and track them to later times. Initially, they have NFW-like density profiles which develop a steeper central slope over time. Using the so far largest axion structure formation simulation with an increased particle number of $2048^3$, we find that MCHs are in better agreement with a modified NFW profile characterized by $\rho\sim r^{-2}$ on small radial scales.
We apply our results to gravitational microlensing, but MCHs with an increased central density will also be of significance for direct and indirect axion detection. 

The structure of this paper is as follows. In \cref{sec:MCs_MCHs} we review the current state of knowledge about axion MCs and MCHs. The procedure of how individual MCs are identified at different simulation snapshots is described in \cref{sec:MC_evolution}, followed by an analysis of the MC evolution. We study the structure of the larger MCHs in \cref{sec:structure_MCH} using the simulations with higher mass resolution. We discuss the implications of our results for axion detection, especially for gravitational microlensing, in \cref{sec:discussion} and we conclude in \cref{sec:conclusion}.

\section{Axion Miniclusters and Minicluster Halos}
\label{sec:MCs_MCHs}

The production of large and spatially extended overdensities $\Phi = (\rho_a-\mean\rho_a)/\mean\rho_a$ in the axion field, where $\rho_a$ denotes the axion energy density and $\mean\rho_a$ its mean value, is characteristic for the post-inflationary symmetry-breaking scenario.
These overdensities decouple from the cosmological expansion at temperature $T = (1+\Phi)T_\mathrm{eq}$, where $T_\mathrm{eq}$ is the temperature at matter-radiation equality, and collapse into gravitationally bound axion MCs~\cite{Hogan1988,Kolb1993,Kolb1994}. The average $\Phi$-dependent density of an MC can be computed from the spherical collapse model and is given by~\cite{Kolb1994}
\begin{align}
    \rho_\mathrm{mc} \simeq 140\Phi^3(1+\Phi)\rho_\mathrm{eq}\,,
    \label{eq:rho_mc}
\end{align}
where $\rho_\mathrm{eq}$ denotes the energy density of the universe at matter-radiation equality. Since the decay of axion strings and topological defects can lead to MC seed overdensities of $\Phi>10^4$ while typical overdensities are of $\mathcal{O}(1)$~\cite{Vaquero2019}, MCs with a broad range in density are formed before matter-radiation equality. Analytical studies predict that the self-similar infall of an isolated density fluctuation results in a radial density profile that can be described by the steep power-law profile~\cite{Bertschinger1985}
\begin{align}
    \rho(r) = \rho_0\left(\frac{r}{r_0}\right)^{-\alpha}
    \label{eq:pl_profile}
\end{align}
with $\alpha=9/4$ and $\rho_0$, $r_0$ are free parameters. As early N-body simulations of the gravitational evolution of the axion field with an initial white-noise power spectrum observed the formation of halos with this predicted power-law profile~\cite{Zurek2006}, \cref{eq:pl_profile} has usually been adopted as the density profile for axion MCs (see for example Refs.~\cite{OHare2017,Fairbairn2017,fairbairn2018}). 

After their formation, the MCs merge hierarchically into larger MCHs~\cite{fairbairn2018,Eggemeier2019_2}. 
It is expected that MCs with a large overdensity survive the merging process with an intact density profile given by \cref{eq:pl_profile} and end up as substructures within the MCHs while less concentrated MCs are tidally disrupted. 
Similar to hierarchical structure formation with CDM, one can predict the average radial density profile of a virialized MCH to agree with the NFW profile~\cite{NFW}
\begin{align}
    \rho(r) = \frac{\rho_s}{(r/r_s)(1+r/r_s)^2}\,,
    \label{eq:nfw_profile}
\end{align}
where $\rho_s$ is the characteristic density and $r_s$ the scale radius. For $r < r_s$ the NFW profile scales as $\rho\sim r^{-1}$ and for $r > r_s$ it is $\rho\sim r^{-3}$, so the scale radius marks the transition between two asymptotic power-law profiles. 

The first N-body simulations studying the gravitational evolution of the highly inhomogeneous axion field from redshift $z=10^6$ to $z=99$ were presented in Ref.~\cite{Eggemeier2019_2}. The results revealed that the first MCs form deep in the radiation-dominated epoch and that the evolution after matter-radiation equality is dominated by halo mergers.
The MCH mass function at $z=99$ is dominated by low-mass halos with masses of $\sim 10^{-15}\,M_\odot$ while the most massive MCHs reach masses of $\sim 10^{-9}\,M_\odot$.\footnote{In principle, halos with even lower masses should also be present but they could not be observed in the simulation due to its limited mass resolution.}
Moreover, the simulation results confirmed that the density profiles of the spatially well-resolved axion MCHs with masses above $10^{-10}\,M_\odot$ can be described by the NFW profile. The scale radius of lower-mass halos could not be spatially resolved but an additional analysis of the density profile of MCHs with masses between $10^{-12}\,M_\odot$ and $10^{-13}\,M_\odot$ showed that it is in agreement with $\rho\sim r^{-3}$~\cite{Eggemeier2022_thesis}, the outer radial slope of the NFW profile. This is further confirmed by the analysis of density profiles at $z=99$ in Ref.~\cite{Ellis2022}. Extending the analysis to the most massive halos at matter-radiation equality, it was found that NFW-like halos are already present at that time while other halos exhibit power-law profiles in agreement with \cref{eq:pl_profile} with $\alpha=9/4$~\cite{Ellis2022, Pierobon2023}.

It might be reasonable to distinguish between MCs and MCHs based on their radial density profile, suggesting that MCs exhibit power-law profiles while MCHs are NFW-like. Such a presumed distinction directly affects estimates for their tidal disruption through stellar interactions as objects with a power-law density profile have a significantly larger survival probability than those with NFW profiles~\cite{Kavanagh2020}. 
For a fixed mass, the power-law profile given by \cref{eq:pl_profile} with $\alpha=9/4$ is more compact than the NFW profile, so the higher survival rate is intuitively comprehensible. 
The density profile of an MC(H) is also crucial for its capability of producing observable microlensing events. It was found that MC(H)s described by an NFW profile are not dense enough and in contrast to the power-law density profile cannot lead to microlensing~\cite{Ellis2022}. Similarly, the expected radio signatures resulting from encounters of axion MC(H)s with neutron stars depend on the assumed density profile~\cite{Edwards2020}. 

\begin{figure}[t]
    \centering
    \includegraphics[width=\columnwidth]{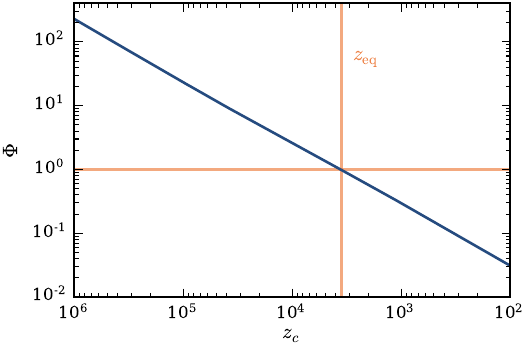}
    \caption{Minicluster seed overdensity $\Phi$ as a function of the collapse redshift $z_c$ of the minicluster. The vertical orange line shows the redshift at matter-radiation equality which translates to an overdensity of $\Phi=1$.}
    \label{fig:Phi_zcoll}
\end{figure}

\section{Tracking Axion Miniclusters}
\label{sec:MC_evolution}

In this section, we continue the analysis of Ref.~\cite{Eggemeier2019_2} and focus on the formation and evolution of axion MCs using the original N-body simulations with $1024^3$ particles. Starting from initial conditions generated by lattice simulations of the early axion field~\cite{Vaquero2019}, the simulations were evolved from $z=10^6$ to $z=99$. Assuming an axion mass of $m_a = 50\,\mu\mathrm{eV}$, the comoving box side length was set to $L=0.864\,\mathrm{pc}$. The comoving softening length that determines the spatial resolution of the simulations was chosen to be $1\,\mathrm{AU}/h$. 
To evolve the Hubble parameter,
\begin{align}
    H(z) = H_0\left(\Omega_{m,0}(1+z)^3+\Omega_{r,0}(1+z)^4+\Omega_{\Lambda,0}\right)^{1/2}
\end{align}
the standard $\Lambda$CDM parameters $\Omega_{m,0}=0.3$, $\Omega_{r,0}=8.486\times 10^{-5}$, $\Omega_{\Lambda,0}=0.7$, and $H_0 = 100h\,\mathrm{km}\,\mathrm{s}^{-1}\mathrm{Mpc}^{-1}$ with $h=0.7$ were used. The \textsc{Subfind} algorithm~\cite{Springel2001,Dolag2009} was applied to identify halos and their subhalos with a minimum number of 32 and 20 gravitationally bound particles, respectively. Halo masses and radii are computed using the virial parameter
\begin{align}
    \Delta_\mathrm{vir}(z) = \frac{18\pi^2 + 82(\Omega_m(z)-1) - 39(\Omega_m(z)-1)^2}{\Omega_m(z)}\,,
    \label{eq:overdens_vir}
\end{align}
where 
\begin{align}
    \Omega_m(z) = \frac{\Omega_{m,0}(1+z)^3}{\Omega_{m,0}(1+z)^3+\Omega_{r,0}(1+z)^4 + \Omega_{\Lambda,0}}\,.
\end{align}
The size of a halo is given by the virial radius $r_\mathrm{vir}$ at which the enclosed mean density coincides with $\Delta_\mathrm{vir}$ times the average matter density $\rho_m = \Omega_m\rho_c$ where $\rho_c=3H^2/(8\pi G)$ is the critical density. Accordingly, the virial mass of a halo is given by $M_\mathrm{vir} = 4\pi/3\Delta_\mathrm{vir}\rho_mr_\mathrm{vir}^3$. 

A conceptual problem with this approach is that it is not directly possible to assign a characteristic density to an identified halo. At any redshift, all halos have the same average density given by $\Delta_\mathrm{vir}\rho_m$. 
This means that the enclosed average density of the very same MC decreases with time, even if its mass stays the same since the physical matter density decreases over time. 
This seems to be in contradiction to \cref{eq:rho_mc} which predicts a characteristic and constant density depending on the MC seed overdensity $\Phi$. However, one can assign to each MC identified by the halo finder its corresponding $\Phi$ value if we assume that its virial density at the redshift $z_c$ of collapse equals \cref{eq:rho_mc}. The result can be seen in \cref{fig:Phi_zcoll} showing that MC seeds with large $\Phi$ values are expected to collapse at $z > z_\mathrm{eq}$ while 
$\Phi=1$ MCs form around matter-radiation equality and objects collapsing at $z<z_\mathrm{eq}$ have $\Phi<1$.
Note that a similar approach of assigning an initial overdensity $\Phi$ to each collapsed MC was already presented in Ref.~\cite{Ellis2022}.

We are interested in the structure of high-density MCs and their evolution. This requires identifying the halos in the snapshots of the N-body simulation that have just collapsed. 
Starting at the high-redshift snapshots, a halo is considered as recently formed if its constituent particles were not gravitationally bound in a halo at a previous snapshot. We then take the corresponding redshift at the given snapshot as the formation redshift of the halo. 
Note that due to the limited number of available simulation snapshots, it is not possible to determine the exact formation redshift of the halos. By construction, all halos collapsed in the time interval between two snapshots are assigned the same formation redshift and thus the same $\Phi$ value.

We register the IDs of the gravitationally bound constituent particles of the recently collapsed MCs as this allows us to locate them again at later redshifts. We then analyze where the high-density MCs forming at $z>z_\mathrm{eq}$ end up at $z=99$ and how their density profiles evolve. 
This offers new insights into MC formation and how they distinguish from the larger MCHs.

\subsection{Minicluster collapse and evolution}
\label{sec:MC_collapse} 

\begin{figure}
    \centering
    \includegraphics[width=\columnwidth]{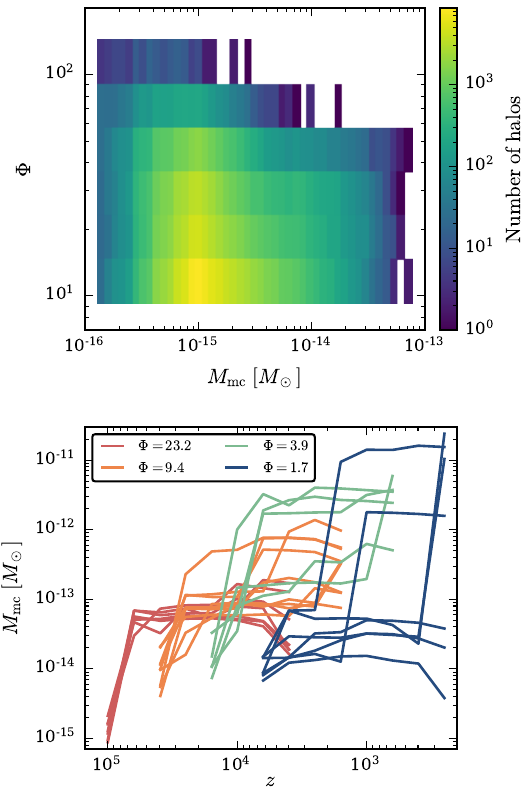}
    \caption{Top: distribution of minicluster overdensity $\Phi$ and mass $M_\mathrm{mc}$ at the time of formation for all objects that have collapsed between $z=6.3\times 10^5$ and $z=3.9\times 10^4$. Note that the visible mass minimum is resolution-dependent and given by the criterion of having at least 32 particles in a halo. Instead of discrete values, the $\Phi$ interval between two consecutive snapshots is considered. Bottom: mass evolution of a sample of miniclusters that are formed at different redshifts and thus have diverse $\Phi$ values (see text for details). This figure uses $1024^3$ particle data.}
    \label{fig:MC_Phi}
\end{figure}

\begin{figure*}
    \centering
    \includegraphics{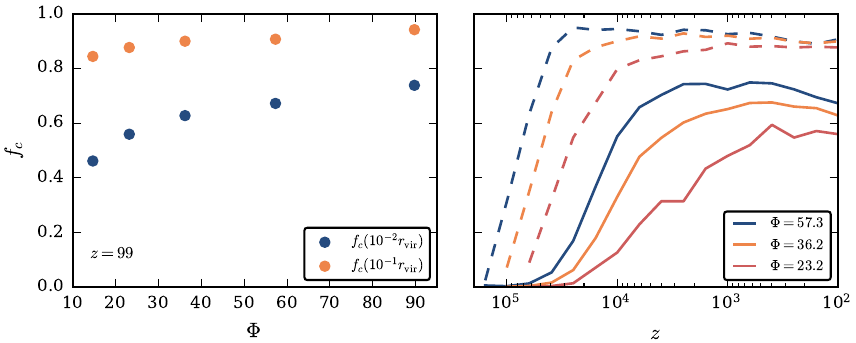}
    \caption{Fraction $f_c$ of minicluster constituent particles that are located within $1\%$ and $10\%$ of the virial radius of the enclosing halo. Left: fraction of minicluster particles contained in the center of larger halos at redshift $z=99$ as a function of $\Phi$ value. Right: redshift evolution of the fraction of minicluster particles located within $1\%$ (solid lines) and  $10\%$ (dashed lines) of the enclosing halo for three different $\Phi$ values. This figure uses $1024^3$ particle data.}
    \label{fig:MC_fraction}
\end{figure*}

The first gravitationally bound objects are identified by the halo finder at redshift $z=3.9\times 10^5$. We note the mass of the halos and the IDs of their constituent particles. We then proceed by determining the recently collapsed halos of the subsequent snapshots. With decreasing redshift our procedure becomes computationally increasingly expensive as more and more halos are identified by the halo finder. Only a fraction of them fulfills our criteria of being recently collapsed and their complete identification for a comprehensive analysis of MCs forming between  $z=3.9\times 10^5$ and matter-radiation equality is not computationally attainable. 
Thus, we focus in the following in general on early forming MCs with high $\Phi$ values but we also track MCs with $\Phi\lesssim 2$.  

The obtained distribution of $\Phi$ and MC mass $M_\mathrm{mc}$ at the time of collapse for all of the emerged objects between $z=6.3\times 10^5$ and $z=3.9\times 10^4$ is shown in the upper panel of \cref{fig:MC_Phi}. 
It is visible that the MCs cover a mass range from $\sim 10^{-16}\,M_\odot$ to $\sim 10^{-13}\,M_\odot$, relatively independent from their corresponding $\Phi$ value. However, the majority of MCs have masses of $\sim 10^{-15}\,M_\odot$. According to the MCH mass distribution (see upper panel of \cref{fig:subhalos_HMF_Phi}), halos with masses larger than $10^{-11}\,M_\odot$ are present at matter-radiation equality. 
Note that the upper panel of \cref{fig:MC_Phi} shows \textit{initial} halo masses while the MCH mass distribution at a certain redshift takes into account not only the just collapsed halos but also the mass increase of previously formed halos. This suggests that halo masses grow significantly via accretion or mergers already during radiation domination and not only after matter-radiation equality, consistent with the findings in Ref.~\cite{Ellis2020}. 

To study the mass evolution of the identified MCs it seems natural to consider the mass computed from the virial overdensity in \cref{eq:overdens_vir} at later times. However, this procedure comes along with a decreasing average density of the MC. Motivated by the spherical collapse model, we instead prefer to keep the MC density given by \cref{eq:rho_mc} fixed. 
To compute the MC mass at later times, we start from its center of mass and integrate radially outwards until the enclosed average density coincides with \cref{eq:rho_mc} with the corresponding $\Phi$ value of the MC. This allows us to determine the MC radius $r_\mathrm{mc}$ independently from virial quantities and we use the enclosed mass within $r_\mathrm{mc}$ as the MC mass.  

Requiring a minimum number of $\mathcal{O}(10^3)$ gravitationally bound particles at the time of formation, we tracked an MC sample with $1.7 \leq \Phi \leq 23.2$ from their formation to the redshift where its radius could still be spatially resolved.\footnote{Axion MCs with $\Phi>23.2$ were not considered in this sample as they consist of significantly less than $\mathcal{O}(10^3)$ particles at their time of formation.} Their mass evolution is shown in the lower panel of \cref{fig:MC_Phi}. 
For MCs with $\Phi\gtrsim 9.4$ one can observe a strong increase in mass initially after their formation that is followed by a mass saturation. Masses ranging from $10^{-13}\,M_\odot$ to $10^{-12}\,M_\odot$ are reached prior to matter-radiation equality, in agreement with expectations~\cite{Eggemeier2019_2}. 
Plenty of surrounding matter may be accreted by these overdense bound structures explaining the fast mass increase. Due to the large time difference between two successive snapshots, we cannot exclude the occurrence of mergers as another explanation attempt.  
A similar behaviour is also visible for some of the $\Phi=3.9$ MCs but a strong initial mass increase does not occur for the MC sample with 
$\Phi\lesssim 1.7$. A substantial mass growth is apparent for some objects of the sample at later times, though, indicating a merger with at least another halo. 

Note that the MC mass saturation also means for the majority of the sample that the corresponding MC radius approaches a constant. 
Hence, the MC radius eventually becomes smaller than the time-increasing physical softening length and it cannot be resolved anymore.
This is in contrast to the virial radius $r_\mathrm{vir}$ of full halo which by definition becomes larger with decreasing redshift for a constant halo mass. 

Following the early forming MCs over time, we can reveal their final location in the simulation box. 
The individual tracking of high-$\Phi$ MCs from our sample suggests that they form the center of larger halos instead of being identified as individual subhalos. 
Based on this observation, we consider the constituent particles of all recently collapsed MCs (the objects from the upper panel of \cref{fig:MC_Phi}) and measure at a given redshift the fraction $f_c$ of them that is located within $1\%$ and $10\%$ of the virial radius of a larger MCH, respectively.

The dependence of the fraction of particles located in the center of MCHs at redshift $z=99$ on the $\Phi$ value of the MCs is shown in \cref{fig:MC_fraction}, together with the redshift-evolution of $f_c$ for MCs with three different $\Phi$ values. 
At redshift $z=99$ a majority of more than $80\%$ of the MC particles, independent of $\Phi$, can be located within $10\%$ of the virial radius of a larger halo. The particle fraction is lower within one percent of the virial radius and there is a stronger dependence on $\Phi$. More particles can be located in the center of larger MCHs for MCs with higher $\Phi$ values and the redshift evolution reveals that the fraction saturates at earlier times for higher values of $\Phi$. This is consistent with the hypothesis that MCs of higher density are more resistant to tidal disruption. 

\begin{figure*}
    \centering
    \includegraphics[width=\textwidth]{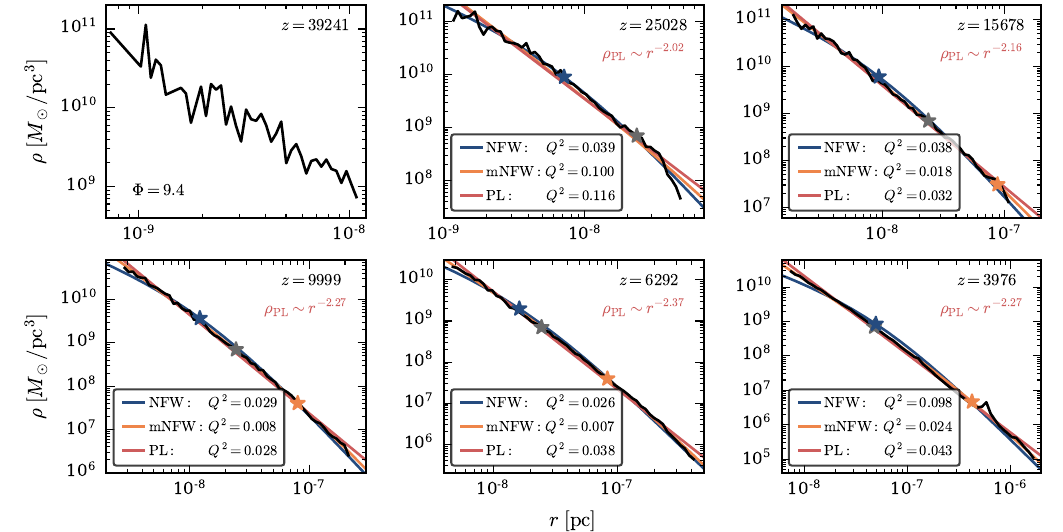}
    \caption{Evolution of the radial density profile of an exemplary minicluster from its formation redshift $z=39241$ to $z=3976$ in physical units. The density profiles are truncated at a radial distance of $4\,\mathrm{AU}/h/(1+z)$ and fitted with the NFW profile from \cref{eq:nfw_profile}, the modified NFW profile from \cref{eq:mod_nfw_profile} and with the power-law (PL) profile from \cref{eq:pl_profile}. The $Q^2$ values (see \cref{eq:Q2}) in the legend measure the quality of the fits with smaller values corresponding to a better fit. The gray star illustrates the radius of the minicluster that is defined to enclose an average density of \cref{eq:rho_mc} with $\Phi=9.4$. The blue and orange stars mark the scale radius of the NFW profile and the modified NFW profile, respectively. This figure uses $1024^3$ particle data.}
    \label{fig:densprofile_mc_evolution}
\end{figure*}

Since the particle fraction analysis is purely statistical, we cannot judge if the particles that originally belonged to an MC still compose an individual gravitationally bound object in the center of the MCH or if they are separated particles.  Related to our finding that the MC radius eventually cannot be resolved anymore, it is understandable that the halo finder cannot identify such dense structures and simply assumes a universal large group of particles in the halos' center. 
On the contrary, it is also possible that the fraction of MC particles that cannot be located in the MCHs' center exist as gravitationally bound subhalos in the exterior of the MCHs.
We continue this discussion in \cref{sec:structure_MCH} where we analyze the MCH substructures identified by the halo finder. 

\begin{figure*}
    \centering
    \includegraphics[width=\textwidth]{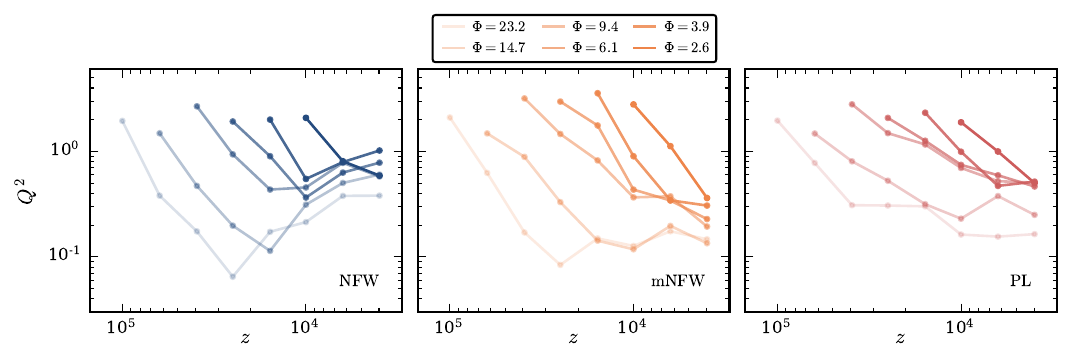}
    \caption{Evolution of average $Q^2$ values (see \cref{eq:Q2}) of miniclusters with $\Phi$ values ranging from $\Phi=23.2$ to $\Phi=2.6$ corresponding to collapse redshifts of $z\simeq10^5$ and $z\simeq 10^4$, respectively. The radial density profiles of the miniclusters were fitted with the NFW profile (left) from \cref{eq:nfw_profile}, the modified NFW profile (center) from \cref{eq:mod_nfw_profile} and with the power-law (PL) profile (right) from \cref{eq:pl_profile} from their collapse redshift until $z=3976$. Smaller values of $Q^2$ correspond to a better fit. This figure uses $1024^3$ particle data.}
    \label{fig:MC_Q2}
\end{figure*}

\subsection{Density profiles of miniclusters}
\label{sec:MC_profiles}

\begin{figure}
    \centering
    \includegraphics[width=\columnwidth]{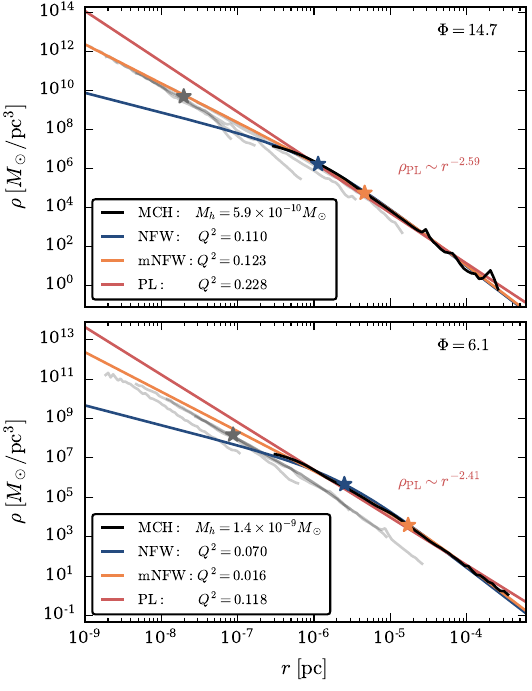}
    \caption{Evolution of the radial density profile of a minicluster with $\Phi=14.7$ (top) and one with $\Phi=6.1$ (bottom). The density profiles at redshifts $z>99$ are shown by the grey curves and the black curve illustrates the density profile at $z=99$ at which time the respective halo exhibits a mass of $M_h = 5.9\times 10^{-9}\,M_\odot$ (top) and $M_h=1.4\times 10^{-9}\,M_\odot$ (bottom). The density profile at $z=99$ is fitted with the NFW profile from \cref{eq:nfw_profile}, the modified NFW profile from \cref{eq:mod_nfw_profile}, and the power-law (PL) profile from \cref{eq:pl_profile} and the best-fit parameterizations are displayed by the blue, orange, and red curves, respectively. The $Q^2$ values (see \cref{eq:Q2}) in the legend measure the quality of the fits with smaller values corresponding to a better fit. The gray star illustrates the radius of the minicluster that is defined to enclose an average density of \cref{eq:rho_mc} with the respective $\Phi$ values. The blue and orange stars mark the scale radius of the NFW profile and the modified NFW profile, respectively. This figure uses $1024^3$ particle data.}
    \label{fig:densprofile_mc_overlayed}
\end{figure}

Tracking individual MCs after their formation we study the evolution of their angular-averaged density profiles. 
In agreement with the resolution studies in Refs.~\cite{Power2002,Zhang2018} and with previous work on MCHs~\cite{Eggemeier2019_2,Ellis2022} we truncate the density profiles at a radial distance of $4\,\mathrm{AU}/h/(1+z)$ in physical units, safely above the numerical softening length. The considered radial distance of the profiles is not limited to the MC radius $r_\mathrm{mc}$ that encloses an average density given by \cref{eq:rho_mc} for a constant $\Phi$ but is determined by the virial radius $r_\mathrm{vir}$ of the entire halo. According to our definition, $r_\mathrm{mc}$ and $r_\mathrm{vir}$ coincide at the time of MC formation but as already mentioned in \cref{sec:MC_collapse} the MC radius eventually approaches a constant while the virial radius continues to increase.  
We fit the density profiles within $r_\mathrm{vir}$ with different models by minimizing the figure-of-merit function~\cite{Navarro2008}
\begin{align}
    Q^2 = \frac{1}{N_\mathrm{bins}}\sum_{i=1}^{N_\mathrm{bins}} (\ln\rho_i - \ln\rho_i^\mathrm{model})^2\,,
    \label{eq:Q2}
\end{align}
where $N_\mathrm{bins}$ denotes the number of logarithmically-spaced bins of the density profile. We use this function to measure the disagreement between the numerical data and the model profiles. Note that $Q^2$ depends only weakly on the bin number whereas the number of free fitting parameters of the model profile is not taken into account at all.

The radial density profile of an exemplary MC with \mbox{$\Phi=9.4$}, corresponding to a collapse redshift of \mbox{$z=39241$}, is shown in \cref{fig:densprofile_mc_evolution} at various times where the MC radius could still be well resolved.
Due to the low number of constituent particles the MC density profile is not yet well defined directly after collapse. 
At the next available snapshot, the scatter is sufficiently reduced which allows us to fit the density profile with the NFW profile from \cref{eq:nfw_profile} and with the single power-law given by \cref{eq:pl_profile} with a free slope parameter $\alpha$. 
At $z=25028$ the MC density profile is in much better agreement with the NFW model than with a single power-law profile. 
This suggests that MCs originating from a direct collapse do not necessarily prefer a single power-law profile as it has been usually assumed. The observed deviations from a pure power-law profile resemble findings from numerical simulations of ultracompact minihalos~\cite{UCMH:gosenca2017} that exhibit NFW profiles if their initial overdensity is slightly non-spherical which was found to be the case for MC seed overdensities~\cite{Vaquero2019}. 

However, the central part of the MC density profile starts to become steeper than the best-fit NFW parameterization at $z=15678$. The deviation between the NFW profile and the numerical data on small scales becomes more pronounced from $z=9999$ until $z=3976$ and the power-law profile provides a comparable or even better fit at those times. 
This apparent ambiguity in the evolution of the density profile motivates us to consider instead an overall profile that can take into account the steeper slope on smaller radial scales. 
Specifically, we introduce the modified NFW profile
\begin{align}
    \rho(r) = \frac{\rho_s}{(r/r_s)^2(1+r/r_s)}\,,
    \label{eq:mod_nfw_profile}
\end{align}
that scales as $\rho\sim r^{-2}$ for $r<r_s$ and as $\rho\sim r^{-3}$ for $r>r_s$. Its best-fit parameterization is also shown in \cref{fig:densprofile_mc_evolution} as the orange curve and from $z=15678$ down to $z=3976$ it provides by far the best description of the MC density profile. 

To verify this finding, we track a sample of MCs with seed overdensities ranging from $\Phi=23.2$ to $\Phi=2.6$ from their formation redshift until $z=3976$ where the MC radius could still be well resolved. Fitting the density profiles with the NFW profile, the modified NFW profile, and a power-law profile we compare the quality of the different fits in \cref{fig:MC_Q2} by computing the average $Q^2$ value for all MCs with the same $\Phi$. 
The MC density profiles are not yet well defined at the time of formation and $Q^2>1$ for the three models. At the next available snapshot after formation, the NFW profile provides on average the best fit. Eventually, the best-fit NFW parameterization cannot match the dense central region (as in \cref{fig:densprofile_mc_evolution}), and the density profiles are better described by the modified NFW profile. Note that the power-law fit performs worse than the best-fit parameterization of the modified NFW profile almost everywhere.

Overall, the studied MCs with $\Phi\in[2.6,\,23.2]$ develop NFW-like profiles soon after their formation but the modified NFW profile becomes the better description with decreasing redshift. 
Combined with the observation from \cref{sec:MC_collapse} that high-density MCs tend to end up in the center of the larger MCHs it is likely that this explains the increase in density on small scales of the overall halo. In other words, mergers of early-forming NFW-like MCs may produce a high-density core within the MCH that is incompatible with the NFW prediction. We are not able to confirm this conjecture directly as resolving possible merger events of early MCs requires a significantly better time resolution of the currently available simulation snapshots. 

Tracking the MC sample from \cref{fig:MC_Q2} to the final redshift $z=99$, their density profiles are again in better agreement with the NFW profile which confirms previous studies of the density profiles of high-mass MCHs~\cite{Eggemeier2019_2,Ellis2022}. 
Suspecting that this is only due to the unresolvable smaller scales, we show the evolution of the density profiles of two exemplary MCs with $\Phi=14.7$ and $\Phi=6.1$, respectively, from their formation redshift to $z=99$ in \cref{fig:densprofile_mc_overlayed}. 
We fit the density profiles of the two high-mass MCHs at $z=99$ which reveals that the original and the modified NFW profile provide a comparable fit according to their $Q^2$ values. 
Interestingly, the best-fit parameterization of the modified NFW profile is in remarkable agreement with the earlier density profiles on the smallest resolved scales in both cases. While the NFW prediction falls short by two orders of magnitude in density compared to the reached density on small scales at early times, the best-fit power-law prediction exceeds the data by two orders of magnitude in density. 
Apart from tidal disruption, there is no reason why the central density should decrease at some point, so it can be expected that the MCHs at $z=99$ are indeed much denser than the NFW prediction on the unresolved scales. This further suggests that the modified NFW profile from \cref{eq:mod_nfw_profile} might be more qualified as a model for axion MCHs and we continue this discussion in \cref{sec:densprofile_MCH}.

\section{Structure of minicluster halos}
\label{sec:structure_MCH}
 
We complement the preceding MC analysis by studying the (sub)structure of the larger axion MCHs. We discuss the evolution of the subhalo mass function and compute the $\Phi$ values for the identified subhalos. Using N-body simulations with an identical set-up as described in \cref{sec:MC_evolution} and in Ref.~\cite{Eggemeier2019_2} but with $2048^3$ particles, a significantly better mass resolution is reached which is beneficial for analyzing the density profiles of MCHs and their subhalos. However, due to computational limitations, the N-body simulations could only be evolved to $z=1584$ and not until $z=99$ as is the case for the $1024^3$ particle simulations.

\subsection{Identified substructures of minicluster halos}

\begin{figure}
    \centering
    \includegraphics[width=\columnwidth]{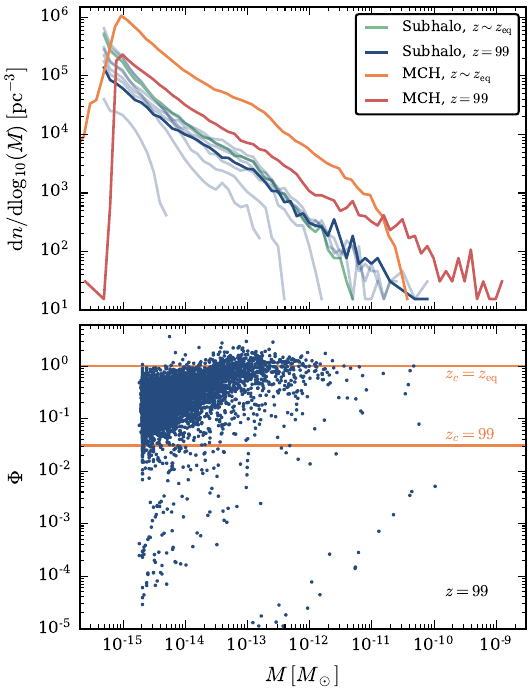}
    \caption{Top: subhalo mass function (light blue colored curves) at different redshifts from $z=1.6\times 10^5$ to $z=99$. The green and dark blue curves mark the subhalo mass function at $z=3976$ close to matter-radiation equality and $z=99$, respectively. The mass functions of the minicluster halos at those times as in Ref.~\cite{Eggemeier2019_2} are shown for comparison. Bottom: distribution of mass and $\Phi$ value for all the subhalos identified at $z=99$ consisting of a minimum number of 100 particles (see text for details). The two orange horizontal lines indicate the expected $\Phi$ for objects collapsing at $z=z_\mathrm{eq}$ and at $z=99$ according to the procedure described in \cref{sec:MC_evolution}. This figure uses $1024^3$ particle data.}
    \label{fig:subhalos_HMF_Phi}
\end{figure}

\begin{figure}
    \centering
    \includegraphics[width=\columnwidth]{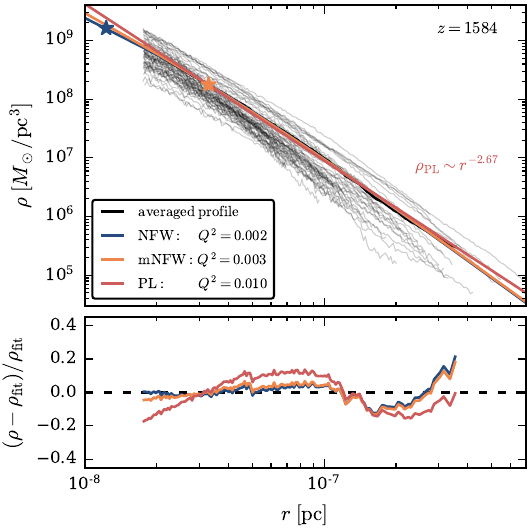}
    \caption{Top: radial density profiles of 50 subhalos with masses between $10^{-13}\,M_\odot$ and $10^{-12}\,M_\odot$ at redshift $z=1584$ in physical units. The density profiles are truncated at a radial distance of $4\,\mathrm{AU}/h/(1+z)$, and the averaged profile is given by the black curve. The blue, orange, and red curves show the best-fit parameterizations of the NFW profile from \cref{eq:nfw_profile}, the modified NFW profile from \cref{eq:mod_nfw_profile}, and the power-law (PL) profile from \cref{eq:pl_profile}, respectively. The $Q^2$ values (see \cref{eq:Q2}) in the legend measure the quality of the fits with smaller values corresponding to a better fit. The blue and orange stars mark the scale radius of the NFW profile and the modified NFW profile, respectively. Bottom: deviations of the averaged density profile from the fits shown in the upper panel. This figure uses $2048^3$ particle data.}
    \label{fig:densprofile_subhalos}
\end{figure}

Collecting the masses of all the subhalos of the MCHs in the $1024^3$ simulations we compute the subhalo mass function that we define as the comoving number density of gravitationally bound subhalos per logarithmic mass interval as a function of subhalo mass. Its evolution from $z=1.6\times 10^5$ to $z=99$ is shown in the upper panel of \cref{fig:subhalos_HMF_Phi}. 
The first subhalos are identified slightly after the first MCs emerge at $z=3.9\times 10^5$, suggesting that mergers occur frequently already in the radiation-dominated era. The subhalos could be MCs that have previously formed or they could originate from remnants of MC mergers. The amplitude of the subhalo mass function increases and subhalo masses of up to $4\times 10^{-12}\,M_\odot$ are reached until matter-radiation equality. 
Afterwards, the overall shape of the subhalo mass function changes barely except that the number density of lower-mass subhalo decreases while higher-mass subhalos of up to $10^{-10}\,M_\odot$ emerge. This could mean that after equality lower-mass subhalos merge to build larger ones, similar to the evolution of MCHs. 

For comparison, we also show the mass function of the MCHs at $z\sim z_\mathrm{eq}$ and $z=99$ in the upper panel of \cref{fig:subhalos_HMF_Phi}. 
The slope of the subhalo mass function in the mass range from $10^{-15}\,M_\odot$ to $10^{-11}\,M_\odot$ is similar to the slope of the MCH mass function at both times confirming previous findings~\cite{Eggemeier2019_2}. Interestingly, the mass range of the subhalos at $z=99$ agrees well with the mass range of MCHs at matter-radiation equality. Although one cannot expect that in general the MCHs from $z_\mathrm{eq}$ end up as subhalos at $z=99$ this is another indication that not only MCs originating from direct collapse during radiation-domination contribute to the substructure of large MCHs. 

To learn more about the roots of the subhalos identified at $z=99$ one can estimate their corresponding overdensity parameter $\Phi$. We are not able to determine their individual time of collapse, so we cannot proceed as done with the MCs in \cref{sec:MC_evolution}. Instead, we trace the subhalos' constituent particles back to the initial conditions and calculate the density of the covered volume. This is done by defining the radius of this seed overdensity as the radius of a spherical shell that is centered at the seed's center of mass enclosing $90\%$ of its total mass. 
Identifying the enclosed density as the subhalo seed density, we use \cref{eq:rho_mc} to obtain the corresponding $\Phi$ value\footnote{We applied this procedure to MCs where the collapse redshift is known and verified that both methods yield similar $\Phi$ values.}. If a subhalo exhibits a value of $\Phi\gtrsim 1$ it can be understood as an MC that collapsed before matter-radiation equality. The origin of subhalos with $\Phi < 1$ is less clear as they can result from a direct collapse at $z<z_\mathrm{eq}$ or from MC(H) merger events. 

\begin{figure*}
    \centering  \includegraphics{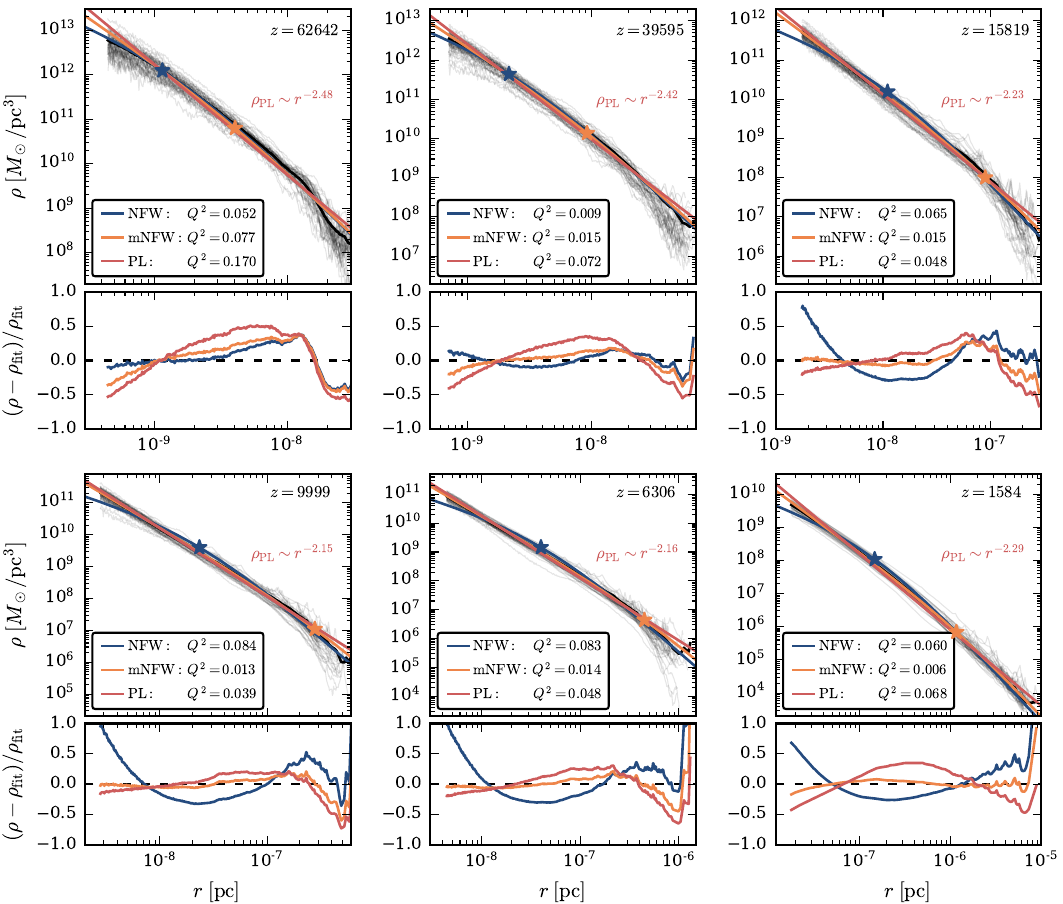}
    \caption{Radial density profiles of the heaviest 50 minicluster halos at different redshifts in physical units. The density profiles are truncated at a radial distance of $4\,\mathrm{AU}/h/(1+z)$, and the averaged profile is given by the black curve. The blue, orange, and red curves show the best-fit parameterizations of the NFW profile from \cref{eq:nfw_profile}, the modified NFW profile from \cref{eq:mod_nfw_profile}, and the power-law (PL) profile from \cref{eq:pl_profile}, respectively. The $Q^2$ values (see \cref{eq:Q2}) in the legend measure the quality of the fits at the respective redshift with smaller values corresponding to a better fit. The blue and orange stars mark the scale radius of the NFW profile and the modified NFW profile, respectively. The deviations of the averaged density profile from the fits are shown in the smaller panels. This figure uses $2048^3$ particle data.}
    \label{fig:MCH_densprofile_evolution}
\end{figure*}

Considering all the subhalos at redshift $z=99$ consisting of at least 100 gravitationally bound particles, we show the distribution of $\Phi$ and subhalo mass in the lower panel of \cref{fig:subhalos_HMF_Phi}. 
It is visible that the majority of the identified subhalos have $\Phi$ values larger than $\Phi(z_c=99)=0.03$ that can be expected for objects collapsing at redshift $z=99$. Subhalos with smaller $\Phi$ values cannot originate from the direct collapse of the seed overdensity, so they have a merger history with progenitor particles that were widely separated in the initial conditions. 
Only a few subhalos were found with $\Phi\geq 1$ which is likely a result of the limited spatial resolution as tracking high-$\Phi$ objects was only possible to redshifts $z>99$ (see \cref{fig:MC_Phi}).
Furthermore, our results are not sufficient to judge if the subhalos in the range $0.03\leq\Phi\leq 1$ are formed from a direct collapse or if they exhibit a relevant merger history. 
Simulations with an increased spatial resolution and a higher number of available snapshots are required for a more detailed analysis of the formation history of the subhalos which we leave to future work.

Using the $2048^3$ particle simulations for an increased mass resolution, we show the radial density profiles of 50 subhalos with masses between $10^{-13}\,M_\odot$ and $10^{-12}\,M_\odot$ at redshift $z=1584$ in \cref{fig:densprofile_subhalos}. 
The halo finder does not determine a subhalo radius, so we define it (in analogy to the procedure of assigning the subhalo $\Phi$) as the radius of a sphere that encloses $90\%$ of the subhalo mass.  
We consider the averaged profile and compare it to the best-fit parameterizations of the NFW profile, the modified NFW profile from \cref{eq:mod_nfw_profile}, and the power-law profile. Both NFW profiles exhibit comparable $Q^2$ values and agree well with the average subhalo profile with a maximum deviation of less than $20\%$. 
However, the scale radius of the pure NFW profile cannot be spatially resolved. 
The best-fit power-law profile matches the data slightly worse suggesting that the subhalo density profiles are NFW-like. Since our spatial resolution is only sufficient to study a rather small radial range, simulations with a considerably better spatial resolution are required for a more detailed analysis.

\subsection{Density profiles of minicluster halos}
\label{sec:densprofile_MCH}

To verify the evidence from \cref{sec:MC_profiles} that the modified NFW profile from \cref{eq:mod_nfw_profile} provides a better description of the structure of axion MCHs, we study their radial density profiles using the $2048^3$ simulations. 
The density profiles of the 50 heaviest MCHs are shown in \cref{fig:MCH_densprofile_evolution} from redshift $z=62642$ to $z=1584$. We compute the average density profile at each redshift and compare it as before to the best-fit parameterizations of the NFW profile from \cref{eq:nfw_profile}, the modified NFW profile from \cref{eq:mod_nfw_profile}, and the power-law profile from \cref{eq:pl_profile}. 
The deviations of the averaged profile from the fits can be seen in the smaller panels. 

It is visible that the density profiles are well described by the NFW profile at early redshifts. However, for $z\lesssim 1.6\times 10^4$ the NFW-fit is not able to capture the high density in the center of the MCHs and the modified NFW profile becomes the better description. This is confirmed by the computed $Q^2$ values (see \cref{eq:Q2}) which are the smallest for the modified NFW profile except for the early redshifts. Furthermore, the averaged MCH density profiles are in remarkable agreement with the modified NFW profile over nearly the entire radial range for the redshifts $z\lesssim 10^4$. Stronger deviations can only be observed in the exterior of the MCHs where the numerical scatter is large. 

Interestingly, the transition at $z\sim 10^4$ from an NFW to the modified NFW profile occurs at a similar redshift when a large fraction of initial MC constituent particles can be located in the center of larger MCHs (see \cref{fig:MC_fraction}). The existence of high-density MCs in the center of MCHs might explain the deviations from the NFW profile on small radial scales and why the slope of the central density profile of MCHs is steeper than previously expected. 

\begin{figure}
    \centering
    \includegraphics[width=\columnwidth]{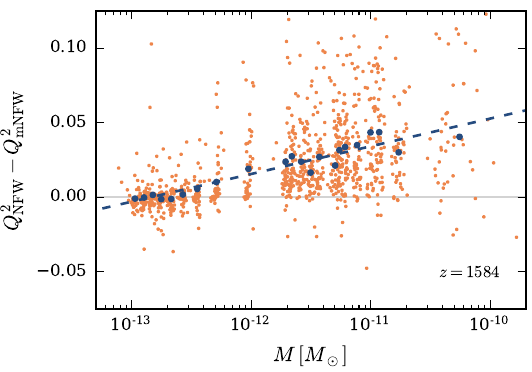}
    \caption{Comparison of the $Q^2$ values (see \cref{eq:Q2}) obtained from fitting a sample of 1100 minicluster halos at redshift $z=1584$ with the NFW (see \cref{eq:nfw_profile}) and with the modified NFW profile (see \cref{eq:mod_nfw_profile}). The orange dots show the difference in $Q^2$ for each halo while the blue dots are averages within different mass ranges. The blue dashed line illustrates the increasing difference of $Q^2$ with increasing MCH mass. Data points above the horizontal line imply that the modified NFW profile is a better fit. This figure uses $2048^3$ particle data.}
    \label{fig:MCH_Q2}
\end{figure}

Since higher-mass MCHs likely draw more high-density MCs into their center, it is conceivable that there might be a mass dependence on which MCH develops a modified NFW profile.
We consider a sample of 1100 MCHs at $z=1584$ with masses ranging from $\sim 10^{-13}\,M_\odot$ to $\sim 10^{-10}\,M_\odot$ and examine their possible mass-dependence on the preferred density profile. The sample is separated into 22 bins consisting of 50 MCHs with similar mass. 
We compare the $Q^2$ values obtained from fits of the MCH density profiles with the original and the modified NFW profile in \cref{fig:MCH_Q2}. If the quantity $Q^2_\mathrm{NFW} - Q^2_\mathrm{mNFW}$ turns negative the NFW profile provides the better fit. Matching the expectations from \cref{fig:MCH_densprofile_evolution} the higher-mass MCHs prefer the modified NFW profile with only a few objects of the sample being better described by the NFW profile. The quality of the two fits converges with decreasing MCH mass and MCHs with masses of $\sim 10^{-13}\,M_\odot$ tend to be better described by the NFW profile. It is noteworthy that also the scatter around the average values reduces significantly with decreasing MCH mass emphasizing the robustness of the NFW profile in the lower-mass regime.  
While we considered only the modified NFW profile with a central slope of $\rho\sim r^{-2}$ it is possible that a profile with a shallower inner slope can also provide an appropriate description, especially for MCHs with intermediate masses of the order of $10^{-12}\,M_\odot$. The dependence of the steepness of the inner slope on the MCH mass deserves a detailed study which we leave to future work. 

\begin{figure}
    \centering
    \includegraphics[width=\columnwidth]{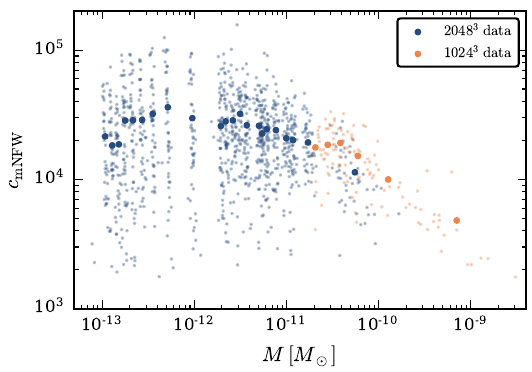}
    \caption{Concentration of minicluster halos extrapolated to $z=0$ (see text for details) obtained from fitting a sample of 1100 minicluster halos at redshift $z=1584$ using the $2048^3$ data and fitting a sample of 110 minicluster halos at redshift $z=99$ using the $1024^3$ data. The larger data points are averaged values. }
    \label{fig:concentration_mass}
\end{figure}

Finally, we use the modified NFW fits to determine the concentration $c_\mathrm{mNFW} = r_\mathrm{vir}/r_s$ of the MCHs. We consider the same sample of 1100 MCHs at redshift \mbox{$z=1584$} as above but we extend it to higher-mass halos by including also 110 MCHs with masses larger than $2\times 10^{-11}\,M_\odot$ from the $1024^3$ particle simulation at redshift $z=99$. As in Ref.~\cite{Ellis2020,Ellis2022}, we evolve the obtained concentrations linearly with scale factor to project our results to redshift $z=0$. The resulting concentration-mass distribution is shown in \cref{fig:concentration_mass} and illustrates that MCHs in the mass range from $10^{-13}\,M_\odot$ to $10^{-10}\,M_\odot$ can be expected to have typical concentrations between $10^4$ and $10^5$ at present time. 
We make use of this distribution in \cref{sec:discussion} to judge if the MCHs are dense enough to give rise to gravitational microlensing.

\section{Phenomenological consequences}
\label{sec:phenomenological_consequences}

Our results from \cref{sec:MC_evolution,sec:structure_MCH} provide evidence that axion MCs and the larger MCHs should not be classified as objects exhibiting single power-law and NFW density profiles, respectively, as it has been usually assumed. Instead, they tend to be in better agreement with the modified NFW profile from \cref{eq:mod_nfw_profile} which produces a steeper inner slope compared to the original NFW profile. This observation might be explained by the presence of high-density MCs in the center of MCHs. 

Direct and indirect axion detection experiments will be affected by MCHs with an increased central density. Studying the stellar disruption of axion MCHs in the Milky Way, it was found that only $46\%$ of MCHs with an NFW profile survive stellar interactions while power-law MCHs have a survival probability of $99\%$~\cite{Kavanagh2020}. We expect an enhanced survival rate for MCHs that are described by the modified NFW profile from \cref{eq:mod_nfw_profile}. 
As a result, fewer tidal streams of axions would be produced making direct axion searches less effective although an encounter with such an MCH would in principle generate a stronger observable signal due to its higher central density. 

In contrast, the likelihood of an indirect detection via gravitational microlensing~\cite{Kolb1995,Fairbairn2017,fairbairn2018} or radio emissions from encounters between MCHs and the magnetospheres of neutron stars~\cite{Tkachev2014,Pshirkov2016,Edwards2020} is raised if more MCHs survive. Estimates of the resulting signal depend in both scenarios crucially on the assumed MCH density profile. 
A study that quantified the properties of the collisions of MCHs with neutron stars found that denser MCHs produce stronger radio signatures~\cite{Edwards2020}. Correspondingly, MCHs with an NFW profile generate weaker signals than power-law MCHs. Moreover, MCHs with an NFW profile cannot lead to observable microlensing signatures in contrast to objects with a power-law profile~\cite{Ellis2022}. One can expect that the modified NFW profile from \cref{eq:mod_nfw_profile} will lead, compared to the original NFW profile, to enhanced signals in both cases. 

In the following, we discuss the microlensing capability of axion MCHs assuming the modified NFW profile. 
The general idea of gravitational lensing is that the path of light of a distant source star is bent by the gravitational field of an object between the star and the observer. This usually leads to multiple images of the source star but in the case of comparatively low-mass axion MCHs separate images cannot be resolved. However, the light of the observed star is amplified when the MCH passes its line of sight which is known as microlensing~\cite{Fairbairn2017,fairbairn2018}. 
For a point-like lens the Einstein radius 
\begin{align}
    R_E = 2\left(GMd(1-d)D_s/c^2\right)^{1/2}
    \label{eq:einstein_radius}
\end{align}
defines the lensing tube that results in an amplification of 1.34~\cite{Fairbairn2017} where $M$ is the lens mass, $D_S$ the distance from the observer to the source, $d = D_L/D_S$ with $D_L$ denoting the distance from the observer to the lens and $c$ is the speed of light. 

\begin{figure}
    \centering
    \includegraphics[width=\columnwidth]{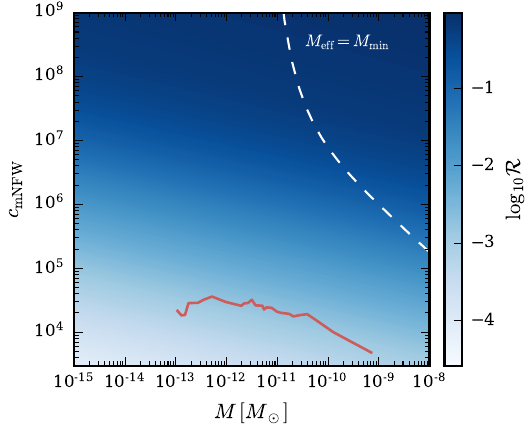}
    \caption{Relative lensing tube size $\mathcal{R}$ for the modified NFW profile from \cref{eq:mod_nfw_profile} as a function of minicluster halo mass $M$ and concentration $c_\mathrm{mNFW}$. The white dashed curve illustrates the required minimum effective lensing mass for an observable microlensing signal (see text for details). The concentration-mass distribution of the simulated minicluster halos obtained from considering the averages in \cref{fig:concentration_mass} is shown for comparison as the red curve.}
    \label{fig:lensing_mNFW}
\end{figure}

For an extended object the lensing tube radius \mbox{$\xi = \mathcal{R}R_E$} is rescaled by a factor $\mathcal{R}\lesssim 1$ and the magnification is given by~\cite{Fairbairn2017,fairbairn2018}
\begin{align}
    \mu = \frac{1}{(1-B)(1+B-C)}\,, 
    \label{eq:lens_magnification}
\end{align}
with parameters
\begin{align}
    C = \frac{1}{\Sigma_c\pi\xi}\frac{\mathrm{d}M(\xi)}{\mathrm{d}\xi}\,,\; B = \frac{M(\xi)}{\Sigma_c\pi\xi^2}\,,\; \Sigma_c = \frac{c^2D_S}{4\pi G D_L D_{LS}}\,,
\end{align}
where $\Sigma_c$ denotes the critical surface mass density and $D_{LS} = D_S-D_L$ is the distance between the lens and the source. The lensing tube is then defined by the value of $\xi$ that results in a magnification of $\mu=1.34$. Following Refs.~\cite{fairbairn2018,Ellis2022}, this corresponds to computing the maximum lensing radius $\xi_\mathrm{max}$ giving a magnification of $\mu=1.17$. The corresponding relative lensing tube size is then $\mathcal{R} = \xi_\mathrm{max}/R_E$. 
This quantity was calculated in Ref.~\cite{Ellis2022} for axion MCHs with an NFW and a power-law profile. We complement this by considering the modified NFW profile from \cref{eq:mod_nfw_profile}.  

We start by computing the surface mass density of the modified NFW profile by integrating \cref{eq:mod_nfw_profile} along the line of sight from $-\infty$ to $\infty$ which yields $\Sigma(x) = 2\rho_s r_s f(x)$ with
\begin{align}
    f(x) = 
    \begin{cases}
        \frac{\pi}{2x} - \frac{2}{\sqrt{x^2-1}}\arctan\sqrt{\frac{x-1}{x+1}} & \mathrm{if}\; x>1\,, \\
        \frac{\pi}{2x} - \frac{2}{\sqrt{1-x^2}}\mathrm{arctanh}\sqrt{\frac{1-x}{x+1}} & \mathrm{if}\; x<1\,, \\
        \frac{\pi}{2}-1 & \mathrm{if}\; x=0\,,
    \end{cases}
\end{align}
where $x=\xi/r_s$ and $\xi$ is the radial coordinate in the lens plane. We then obtain the surface mass profile that appears in \cref{eq:lens_magnification} by solving the integral
\begin{align}
    M(\xi) = 2\pi\int_0^\xi\Sigma(\xi^\prime/r_s)\xi^\prime\mathrm{d}\xi^\prime
\end{align}
numerically. With this, it is possible to solve \cref{eq:lens_magnification} for $\xi_\mathrm{max}$ as a function of MCH mass and concentration parameter $c_\mathrm{mNFW} = r_\mathrm{vir}/r_s$. Following Refs.~\cite{fairbairn2018,Ellis2022}, we consider the distance to the Andromeda galaxy as the distance to the source and we average over different lens positions to obtain an average value for $R_E$ and $\Sigma_c$. The resulting distribution of $\mathcal{R}$ is shown in \cref{fig:lensing_mNFW}. There exists an absolute survey-dependent mass limit below which a lens is not able to generate a microlensing signal. For the Subaru HSC survey, this mass limit is $M_\mathrm{min}\approx 3\times 10^{-12}\,M_\odot$~\cite{Niikura2017}. The effective lensing mass $M_\mathrm{eff}$ enclosed by $\xi_\mathrm{max}$ needs to exceed $M_\mathrm{min}$ to generate an observable signal. This strict boundary is visualized in \cref{fig:lensing_mNFW} revealing that microlensing with MCHs can be only observed for high-mass objects with large concentrations. 

In contrast to microlensing of MCHs with an NFW profile~\cite{Ellis2022} a sharp boundary between $\mathcal{R}=0$ and $\mathcal{R}>0$ originating from the shallow $\rho\sim r^{-1}$ inner slope does not exist for the modified NFW profile. Moreover, compared to the NFW predictions lower concentrations are already sufficient to fulfill the lensing mass criterion. 
The obtained concentration-mass distribution from \cref{fig:concentration_mass} suggests that the simulated MCHs have too low concentrations to give rise to gravitational microlensing.
However, unresolved mergers of dense MCs might lead to an overall much denser central MCH region which would shift the effective lensing mass closer to the required threshold of $M_\mathrm{min}$. Only simulations with a better spatial resolution will be able to reveal if this is the case.  

\section{Discussion}
\label{sec:discussion}
Structure formation with axion dark matter in the post-inflationary symmetry-breaking scenario is characterized by the formation of dense axion miniclusters~(MCs) that collapse before matter-radiation equality and merge hierarchically into larger axion minicluster halos~(MCHs). 
It is important to study the MC(H) structure and how MCs evolve into MCHs to assess their phenomenological implications. 
In previous work, the density profiles of higher-mass MCHs were resolved at $z=99$ and at matter-radiation equality but the evolution of the density profiles starting from the formation redshift was not considered at all~\cite{Eggemeier2019_2,Ellis2022,Pierobon2023}. So it remained for example unclear if MCs exhibit, as suggested by theoretical arguments (see \cref{sec:MCs_MCHs}), power-law density profiles right after their formation and how they evolve. 
Using the $1024^3$ N-body simulations from Ref.~\cite{Eggemeier2019_2} we have followed the evolution of MCs from their formation redshift to $z=99$. We studied their density profiles and made use of N-body simulations with $2048^3$ particles for a detailed analysis of the structure of the larger MCHs. 

Identifying objects that have recently collapsed at different simulation snapshots we collected a sample of MCs and assigned them a characteristic seed overdensity $\Phi$ (see \cref{eq:rho_mc}). This allowed us to track the MCs to later times and to observe how their evolution depends on $\Phi$. Covering a mass range of $\sim 10^{-16}\,M_\odot$ to $\sim 10^{-13}\,M_\odot$ at the time of their formation independent of their respective $\Phi$ value, a higher fraction of MC constituent particles ends up in the center of larger MCHs at $z=99$ for objects with higher $\Phi$. This is comprehensible as denser MCs should be more resistant to tidal disruption. 

While tracing the MCs to lower redshifts, we observed that the radial density profile of an isolated MC is initially well described by the NFW profile rather than by a single power-law profile. During the evolution, the MCs became part of larger MCHs with an overall steeper inner slope. This motivated us to introduce the modified NFW profile from \cref{eq:mod_nfw_profile} for MCHs, where the density of the inner region approaches $\rho\sim r^{-2}$ instead of $\rho\sim r^{-1}$ for the NFW profile.
Eventually, the overall density profiles turned out to be in better agreement with the modified NFW profile which is even able to match the early MC density profiles on the smallest resolved scales.

We complemented the MC analysis by examining the structure of the larger MCHs and their identified subhalos and we confirmed that the subhalo mass function is similar to the overall MCH mass function.
Aiming to reveal if the subhalos originate from a direct collapse during radiation domination we traced their constituent particles back to the initial conditions and computed their seed overdensity $\Phi$. 
Most of the subhalos exhibit a seed overdensity in the range $0.03\leq\Phi\leq 1$ which corresponds to collapse redshifts between $z_\mathrm{eq}$ and $z=99$.
The fact that subhalos with $\Phi$ values comparable to those of the densest MCs were not identified agrees with our previous result that the densest MCs tend to end up in the center of MCHs and not as gravitationally bound subhalos. This is likely due to limitations in the spatial resolution as the halo finder might not be able to resolve the dense MCs in the center of the MCHs as individual subhalos. 
Using the $2048^3$ particle simulations we also analyzed the subhalo density profiles at redshift $z=1584$ in the mass range from $10^{-13}\,M_\odot$ to $10^{-12}\,M_\odot$ and found that they are NFW-like. However, simulations with a better spatial resolution are required for a more extensive study.

Inspired by the discovery that the existence of dense MCs in the center of MCHs seems to affect the inner slope of the MCH density profile, we studied the evolution of the higher-mass MCHs in the $2048^3$ particle simulations. 
Matching our expectations, the density profiles agree initially with the NFW profile but from $z\sim 10^4$ onwards the modified NFW profile from \cref{eq:mod_nfw_profile} provides a better description. 
We further investigated if this is also true for lower-mass MCHs and observed that the quality of the modified NFW fit decreases with decreasing halo mass.In particular, MCHs with masses of $\sim 10^{-13}\,M_\odot$ do not have a steeper inner slope and they are thus in better agreement with the original NFW profile. Our results provide evidence that MCHs are described by a class of NFW-like density profiles with varying steepness of the inner slope depending on the halo mass. This is in contrast to the previous perception of MCHs having NFW profiles and MCs with power-law profiles.

Axion MCHs with a higher central density than predicted by the NFW profile have important implications for direct and indirect axion detection experiments. 
They are likely more robust to tidal disruption which decreases the probability of direct but increases the likelihood of indirect detection. 
The microlensing signal generated by MCHs with a modified NFW profile is in contrast to the signal from purely NFW halos potentially observable. However, the MCHs from our simulations are not dense enough to be detected by microlensing (see \cref{fig:lensing_mNFW}).
It is expected that denser MCHs in general  will also produce stronger radio signals in encounters with neutron stars. 
Hence, a sensible next step is to estimate the survival probability of MCHs with the modified NFW profile and the radio signature from neutron star encounters as done in Refs.~\cite{Kavanagh2020,Edwards2020}. 

Simulations capable of resolving early MC mergers should be performed to study in detail how mergers affect the central region of the MC density profile. This could also clarify if MCs with a higher $\Phi$ value develop a steeper inner slope and could be used to track the evolution of the objects that end up as subhalos in MCHs. Independently, it should be validated if similar characteristics in the evolution of MC(H)s that were observed in this work can also be confirmed in simulations starting from different initial conditions. 
A recent study suggests that the mass distribution and internal structure of the MCHs seem to depend to some degree on the simulation techniques used to compute the dark matter axion field from axion strings~\cite{Pierobon2023}. It should thus be excluded that our findings are an artifact of how the initial conditions were generated.

\section{Conclusions}
\label{sec:conclusion}

Building on previous studies~\cite{Eggemeier2019_2,Ellis2020,Ellis2022} we used N-body simulations to resolve the evolution of early forming axion miniclusters~(MCs) into larger axion minicluster halos (MCHs). 
Identifying MCs with high overdensities of $\Phi \gg 1$ in the $1024^3$ N-body simulations of Ref.~\cite{Eggemeier2019_2} at the time of collapse and tracing them to lower redshifts, we observe that a large fraction of them ends up near the center of larger MCHs at late times. 
In contrast to theoretical expectations predicting a single power-law profile for objects forming from the self-similar infall of an isolated density fluctuation~\cite{Bertschinger1985}, we find that axion MCs exhibit an NFW profile soon after their formation. 
However, the MC density profile evolution is not finalized at this stage and the MCs subsequently develop a steeper slope approaching $\rho\sim r^{-2}$ on small radial scales.  

The evidence for MCs with an increased central density motivated us to perform an N-body simulation with a higher mass resolution of $2048^3$ particles.
Starting from the same initial conditions as the $1024^3$ particle simulation, we were able to evolve the $2048^3$ simulation to a final redshift of $z=1584$.  
Analyzing the density profile evolution of MCHs we observe a similar behaviour as in the $1024^3$ simulation. Specifically, a modified NFW profile with an inner slope of $\rho\sim r^{-2}$ provides a better fit than the original NFW profile for MCHs with masses above $\sim 10^{-12}M_\odot$ at $z=1584$ whereas the NFW profile well describes the lighter MCHs and MCs. 
Overall, our findings are in contrast to the previous assumptions of MCHs exhibiting NFW profiles and MCs having power-law density profiles~\cite{fairbairn2018,Kavanagh2020,Edwards2020}.  

Considering axion MCHs with the modified NFW from \cref{eq:mod_nfw_profile} we followed Refs.~\cite{Fairbairn2017,fairbairn2018,Ellis2022} and computed the expected signal from gravitational microlensing. In principle, MCHs with an inner slope of $\rho\sim r^{-2}$ are in contrast to pure NFW halos dense enough to generate an observable signal but the concentrations of the MCHs from our simulations are still too small.
In general, more compact MCHs will be more robust to tidal disruption which decreases the likelihood of direct detection. Conversely, one can expect stronger signals in indirect detection experiments such as encounters of MCHs with neutron stars.  

It is therefore important to study the density profile of MCHs and in particular their central region in great detail. Our findings provide new insights into the structure of axion MCHs with evidence for an increased central density. However, there might be some dependency of the MCH density profile on the simulation technique that is used to generate the initial conditions of our N-body simulations. Hence, it needs to be validated that our results in this work can be also confirmed in simulations starting from different initial conditions.

\section*{Acknowledgements}

We thank David Ellis, Doddy Marsh, Jens Niemeyer, Giovanni Pierobon, and Javier Redondo for valuable discussions. 
BE gratefully acknowledges the computing time provided on the supercomputers Lise and Emmy at NHR@ZIB and NHR@Göttingen as part of the NHR infrastructure. This work also used the Scientific Compute Cluster at the GWDG, the joint data center of the Max Planck Society for the Advancement of Science (MPG), and the University of Göttingen.
KD acknowledges support by the COMPLEX project from the European Research Council (ERC) under the European Union’s Horizon 2020 
research and innovation program grant agreement ERC-2019-AdG 882679 as well as support by the Deutsche Forschungsgemeinschaft (DFG, 
German Research Foundation) under Germany’s Excellence Strategy - EXC-2094 - 390783311.
Finally, we acknowledge the use of the open-source Python scientific computing packages NumPy~\cite{numpy}, SciPy~\cite{scipy}, and Matplotlib~\cite{matplotlib}.

\bibliography{refs}

\end{document}